\documentclass[%
 aip,
 jmp,%
 amsmath,amssymb,
 reprint,%
]{revtex4-1}

\usepackage{graphicx}
\usepackage{dcolumn}
\usepackage{bm}
\usepackage{upgreek}
\usepackage{color}

\begin{document}

\preprint{AIP/123-QED}

\title{Infrared spectroscopy of molecular ions in selected rotational and spin-orbit states}

\author{U. Jacovella}
\affiliation{Laboratorium f\"ur Physikalische Chemie, ETH Z\"urich, CH-8093 Zurich, Switzerland}
\author{J. A. Agner}
\affiliation{Laboratorium f\"ur Physikalische Chemie, ETH Z\"urich, CH-8093 Zurich, Switzerland}
\author{H. Schmutz}
\affiliation{Laboratorium f\"ur Physikalische Chemie, ETH Z\"urich, CH-8093 Zurich, Switzerland}
\author{J. Deiglmayr}
\affiliation{Laboratorium f\"ur Physikalische Chemie, ETH Z\"urich, CH-8093 Zurich, Switzerland}
\author{F. Merkt}
\email{frederic.merkt@phys.chem.ethz.ch}
\affiliation{Laboratorium f\"ur Physikalische Chemie, ETH Z\"urich, CH-8093 Zurich, Switzerland}

\date{\today}

\begin{abstract}
First results are presented obtained with an experimental setup developed to record IR spectra of rotationally state-selected ions. The method we use is a state-selective version of a method developed by S. Schlemmer, D. Gerlich and coworkers (Int. J. Mass. Spec. \textbf{185}, 589 (1999); J. Chem. Phys. \textbf{117}, 2068 (2002)) to record IR spectra of ions. Ions are produced in specific rotational levels using mass-analyzed-threshold-ionization spectroscopy. The state-selected ions generated by pulsed-field ionization of Rydberg states of high principal quantum number ($n\approx200$) are extracted toward an octupole ion guide containing a neutral target gas. Prior to entering the octupole, the ions are excited by an IR laser. The target gas is chosen so that only excited ions react to form product ions. These product ions are detected mass selectively as a function of the IR laser wavenumber. To illustrate this method, we present IR spectra of C$_2$H$_2^+$ in selected rotational levels of the $^2\Pi_{\textrm{u,}3/2}$ and $^2\Pi_{\textrm{u,}1/2}$ spin-orbit components of the vibronic ground state.
\end{abstract}

\pacs{Valid PACS appear here}
\maketitle


\section{Introduction}
Many important molecular cations such as H$_3^+$, CH$_4^+$, CH$_5^+$, C$_2$H$_4^+$, C$_2$H$_5^+$, C$_2$H$_6^+$, C$_5$H$_5^+$, C$_6$H$_6^+$, O$_3^+$, etc., have complex or unknown spectra. The presence of unpaired electrons, orbital degeneracy and$/$or large-amplitude motions in these ions leads to intricate energy-level structures and congested spectra. Once measured, the spectra can resist assignments for years, if not for decades. Striking examples include the Carrington-Buttenshaw-Kennedy spectrum of H$_3^+$,\cite{Carrington1982} the White-Tang-Oka spectrum of CH$_5^+$,\cite{White1999} and the spectrum of CH$_4^+$.\cite{signorell1999,woerner2009}

The difficulties in studying cations arise from their high reactivity and the fact that space-charge effects limit their densities in the gas phase. In most cases, ions are generated in hot environments so that the population is distributed over many quantum states, which further reduces the sensitivity of spectroscopic experiments and complicates spectral assignments.

Several approaches have been developed to overcome these difficulties. These can be classified into three categories: i) the direct measurement of ion spectra  by absorption or emission spectroscopy, \cite{Herzberg1971,Saykally1981,Hirota1992,Hirota2000} the sensitivity of which is greatly enhanced by exploiting modulation techniques,\cite{Oka2003,Stephenson2005,Hodges2013} ii) the indirect measurement of ion spectra by photoelectron spectroscopy of the neutral parent molecules, for instance by pulsed-field-ionization zero-kinetic-energy (PFI-ZEKE) photoelectron spectroscopy, \cite{Muller1991,Merkt2011} and iii) the measurement of spectra by detecting photoinduced processes, \emph{e.g.} dissociation products,\cite{Dzhonson2006,Rizzo2015} reaction products,\cite{Schlemmer1999,Schlemmer2002} the suppression of complex formation,\cite{Duncan2003,Chakrabarty2013} or specific processes in ion traps resulting from photoexcitation~\cite{Tong2011}. A variant of iii) consists of recording the spectra of weakly bound complexes of the ion of interest with rare gas atoms and monitoring the rare-gas-atom loss.\cite{Duncan2011} Significant advances could be made in this last category with the introduction of 22-pole ion traps in combination with buffer-gas cooling~\cite{Gerlich1993} and the use of sympathetically cooled molecular ions in laser-cooled Coulomb crystals.\cite{Willitsch2012} Several recent review articles provide an overview of this very active field.\cite{Willitsch2011,Amano2011}

We present here a method for recording spectra of \textit{state-selected} ions and illustrate it with the example of a measurement of the asymmetric-stretch fundamental band (3$_0^1$) of C$_2$H$_2^+$. The transitions are detected by monitoring the C$_2$H$_3^+$ product of the reaction
\begin{equation}
\textrm{C}_2\textrm{H}_2^+ \textrm{(} 3^1, \Omega^+, J^+ \textrm{)}+ \textrm{H}_2 \rightarrow \textrm{C}_2\textrm{H}_3^+ + \textrm{H}
\label{Equation1}
\end{equation}
which is not observable for C$_2$H$_2^+$ ions in their vibronic ground state, as was already exploited in the pioneering experiment of Schlemmer $et$ $al.$\cite{Schlemmer2002}. In Eq.~(\ref{Equation1}), $J^+$ is the total-angular-momentum quantum number excluding nuclear spin and $\Omega^+$ is the quantum number associated with the projection of $J^+$ on the molecular axis. C$_2$H$_2^+$ is particularly well suited to test the performance of our method because the transition frequencies are accurately known from the work of Oka and coworkers \cite{Jagod1992} and the spin-orbit and rotational structure of the photoelectron spectrum of the X$^+$ $\leftarrow$ X ionizing transition has been fully resolved by PFI-ZEKE photoelectron spectroscopy \cite{Rupper2004}.

Carrying out the experiments with state-selected ions makes the spectral assignments straightforward, reduces spectral congestion and allows in principle the measurement of infrared spectra of vibrationally excited ions.

\section{Experimental procedure}
\begin{figure*}[!ht]\centering
\includegraphics[width=1\linewidth]{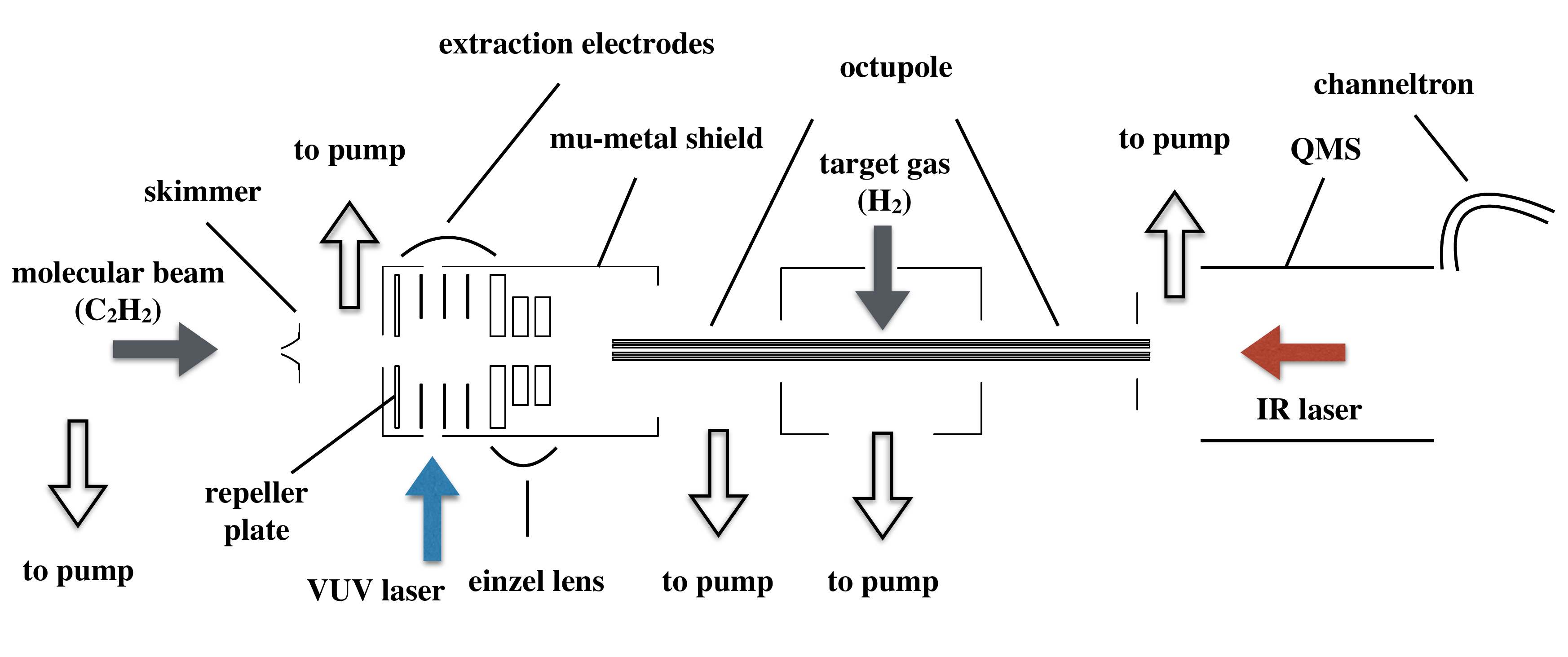}
\caption{Schematic diagram of the experimental setup. Different components are not to scale. See text for explanations.} 
\label{Figure1}
\end{figure*}

The first step of our experimental procedure consists in preparing the C$_2$H$_2^+$ cations in selected rotational levels of the ground electronic state and generating a beam of these ions. This step is accomplished using mass-analyzed-threshold-ionization (MATI)~\cite{Zhu1991,Merkt1993} combined with single-photon excitation of a cold sample of C$_2$H$_2$ molecules in a supersonic beam using narrow-band vacuum-ultraviolet (VUV) laser radiation.

The C$_2$H$_2^+$ ions are irradiated with an IR laser pulse before being extracted toward an octupole ion guide containing the neutral H$_2$ target gas. The IR transitions are detected by monitoring the C$_2$H$_3^+$ product-ion yield as a function of the IR-laser wavenumber. The experimental setup is represented schematically in Fig.~\ref{Figure1}.

The pulsed tunable VUV laser radiation (repetition rate 25~Hz, pulse duration 2~ns) was generated by resonance-enhanced difference-frequency mixing in a pulsed beam of xenon~\cite{Rupper2004} using the (5p)$^6$ $\rightarrow$ (5p)$^5$6p[1/2]$_0$ two-photon resonance of Xe at $2\tilde\nu_{1}= 80118.98$\,cm$^{-1}$ and two commercial single-grating dye lasers and a $\beta$-barium-borate doubling crystal. 

The pulsed tunable IR laser radiation was generated by difference-frequency mixing in a KTiOAsO$_4$ (KTA) crystal. The KTA was cut at angles of $\Theta=43^{\circ}$, $\Phi=0^{\circ}$ and had a dimension of $5\times5\times15$\,mm$^3$. The seeded 532\,nm output of the Nd:YAG laser and the output of a double-grating tunable dye laser operated in the range 638-642\,nm were overlapped using a dichroic mirror and directed into the crystal. To separate the generated IR laser beam from the input laser beams, these were aligned so as to intersect in the KTA crystal at a small angle ($<5^\circ$). A slit was used to block the 532\,nm and 640\,nm laser beams but transmit the IR laser beam before the IR beam entered the experimental chamber through a CaF$_2$ window. The IR radiation had a measured bandwidth of 0.035\,cm$^{-1}$ and a pulse energy of 200\,$\upmu$J. Its wavenumber was calibrated at an absolute accuracy of 0.02\,cm$^{-1}$ by measuring the wavenumber of the 532\,nm and 640\,nm lasers with a wavelength meter.

A pulsed valve was employed to form a supersonic beam of a mixture of 20$\%$ acetylene and 80$\%$  argon (volume $\%$). The beam was skimmed prior to entering the photoionization chamber, where it intersected the VUV laser beam at right angles on the axis of a set of five parallel cylindrical extraction plates.\cite{Rupper2004}

The ions were state selected using two successive electric-field pulses,\cite{Merkt1993} a discrimination pulse of +690\,mV/cm, which removed all prompt ions produced by the VUV laser, followed by a field-ionization and ion-extraction pulse of $+$5.5\,V/cm, generated by applying a pulsed electric potential of +32\,V on the repeller plate (see Fig.~\ref{Figure1}) of the extraction region. The IR excitation was carried out either during the discrimination pulse or in the field-free interval between discrimination and extraction pulses, or immediately following field ionization with the extraction pulse. In this way the Doppler shift of the IR transition only originated from the velocity of the supersonic beam ($\approx$600 m/s), and was $\approx$ $-$0.006~cm$^{-1}$ at 3000~cm$^{-1}$, i.e., more than $\approx$ 5 times less than the IR-laser bandwidth. The IR spectra turned out not to be sensitive to the exact timing of the IR laser pulse as long as the IR excitation took place before significant acceleration of C$_2$H$_2^+$ by the extraction pulse. We therefore conclude that the presence of the Rydberg electron has no effect on IR transition frequencies within the precision limit of our experiment and of electric fields up to $+$5.5~V/cm. However, we observed a slight broadening (from 0.035~cm$^{-1}$ to 0.050~cm$^{-1}$) of the IR transitions recorded before the field-ionization pulse (see below).

The ions extracted toward the reaction zone were produced at a potential of $\approx18$\,V, determined by the applied potential and the experimental geometry, and were slowed down on their way to the octupole containing the target gas H$_2$ by applying a $\approx17$\,V bias potential to the octupole rods. The exact value of the bias potential was carefully adjusted so that i) the kinetic energy of the ions entering the reaction zone was too low for a reaction to occur without prior excitation of C$_2$H$_2^+$ with the IR laser, and ii) the prompt ions were all rejected.

\begin{figure}[!ht]\centering
\includegraphics[width=1\linewidth]{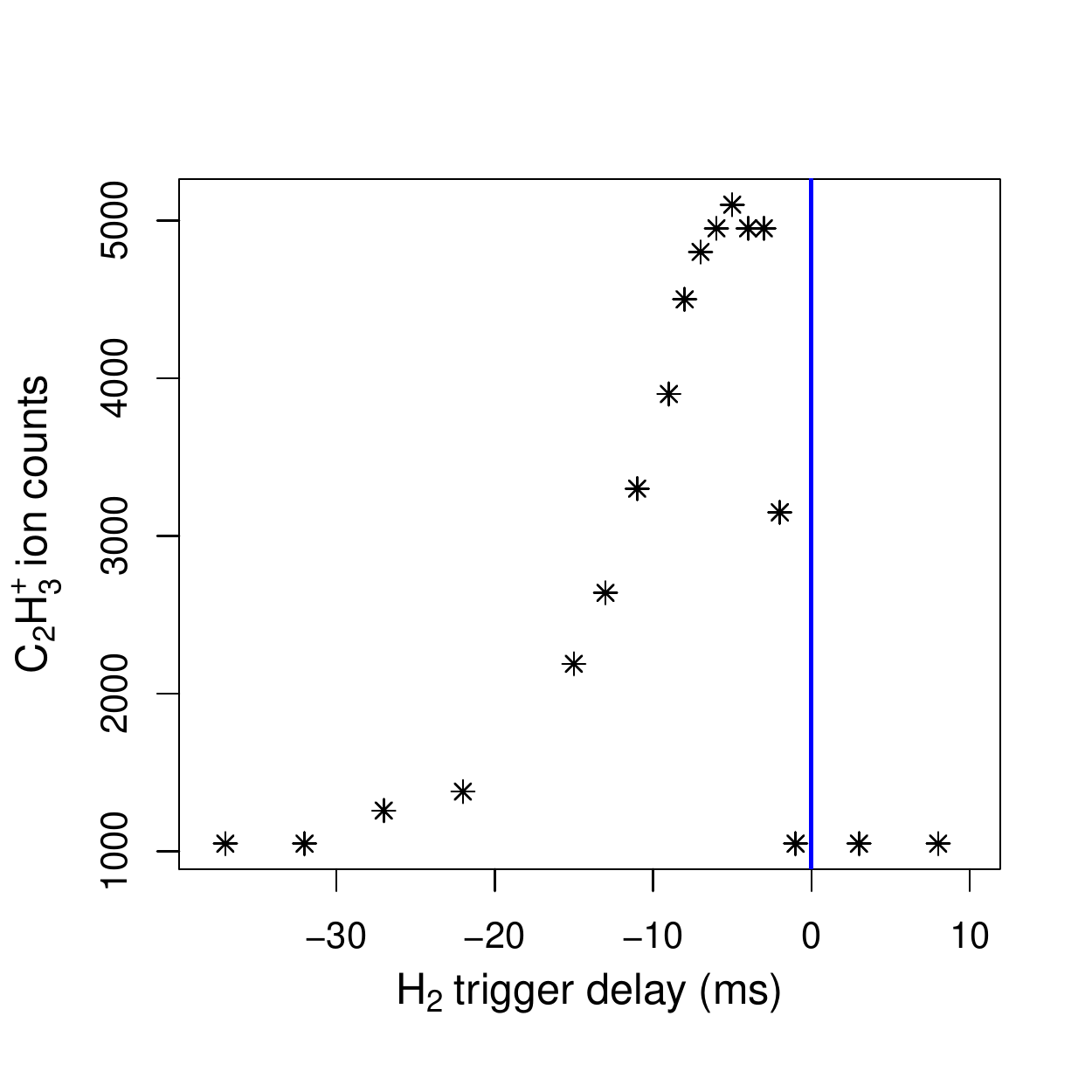}
\caption{\textcolor{red}{C$_2$H$_3^+$ product yield (obtained from the average of 600 experimental cycles) as a function of the H$_2$ valve opening trigger time with respect to the VUV laser pulse (vertical line). To induce the reaction for this measurement, the C$_2$H$_2^+$ ions were accelerated to a kinetic energy sufficient to overcome the barrier without IR laser.}} 
\label{Figure2}
\end{figure}

The target gas H$_2$ was introduced from the side using a pulsed valve with orifice located 10\,cm away from the octupole axis. Rather than monitoring the product ions resulting from the reaction of C$_2$H$_2^+$ with H$_2$ molecules in the supersonic beam, we found it more efficient to open the H$_2$ valve early and to let the H$_2$ molecules thermalize to 300\,K prior to the arrival of the C$_2$H$_2^+$ beam in the octupole. The optimal valve-opening time was determined by monitoring the C$_2$H$_3^+$ product-ion yield, as illustrated in Fig.~\ref{Figure2}. In this figure, the vertical line indicates the temporal position of the VUV laser pulse and the stars represent the C$_2$H$_3^+$ ion signal. The maximum yield was obtained by opening the H$_2$ valve about 6~ms before the VUV laser pulse. After the reaction, the product ions (C$_2$H$_3^+$) were guided toward a commercial quadrupole mass spectrometer coupled with an off-axis channeltron, where they were detected mass-selectively.   

\section{Results}

All assignments are presented using Hund's case (a) nomenclature, which is adequate for low values of $J^+$. The IR transitions are labeled R$_{1,2}$($J^+$), where R is the standard notation for transitions associated with $\Delta J^+$ =+1, $J^+$ is the rotational quantum number of the initial state and the subscripts 1 and 2 refer to the F$_1$ ($^2\Pi_{3/2}$) and F$_2$ ($^2\Pi_{1/2}$) spin-orbit components of the X$^+$ ground electronic state of C$_2$H$_2^+$.\\

\begin{figure}[!ht]\centering
\includegraphics[width=1\linewidth]{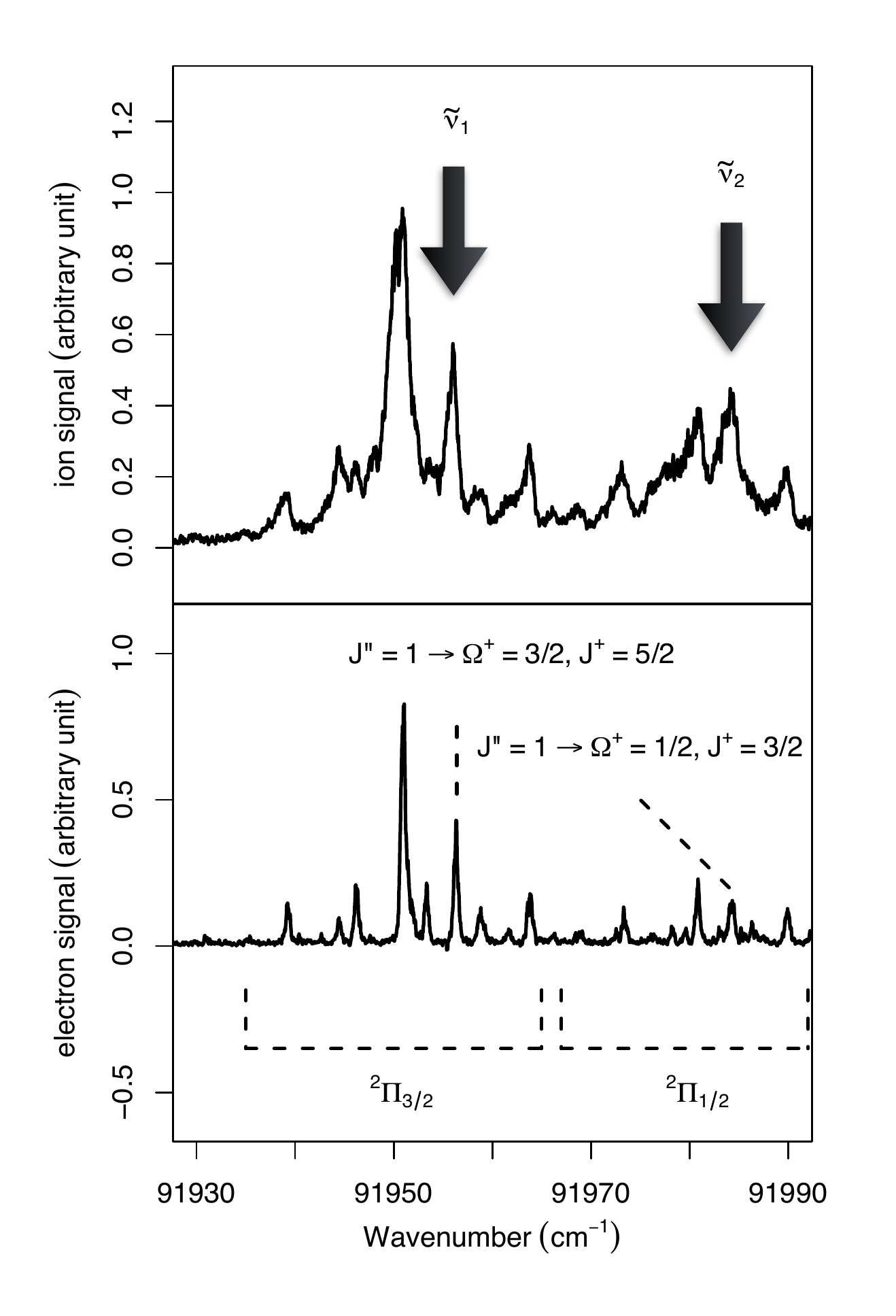}
\caption{Comparison of the MATI spectrum (upper trace) and PFI-ZEKE photoelectron spectrum (lower trace, from Ref. [\onlinecite{Rupper2004}]) of the X$^+$ $^2\Pi_\textrm{u}$ ($v^+$=0) $\leftarrow$ X $^1\Sigma_\textrm{g}$ ($v$=0) ionizing transition of C$_2$H$_2$. Both spectra have been corrected for the field-induced shift of the ionization thresholds.} 
\label{Figure3}
\end{figure}

\begin{figure}[!ht]\centering
\includegraphics[width=1\linewidth]{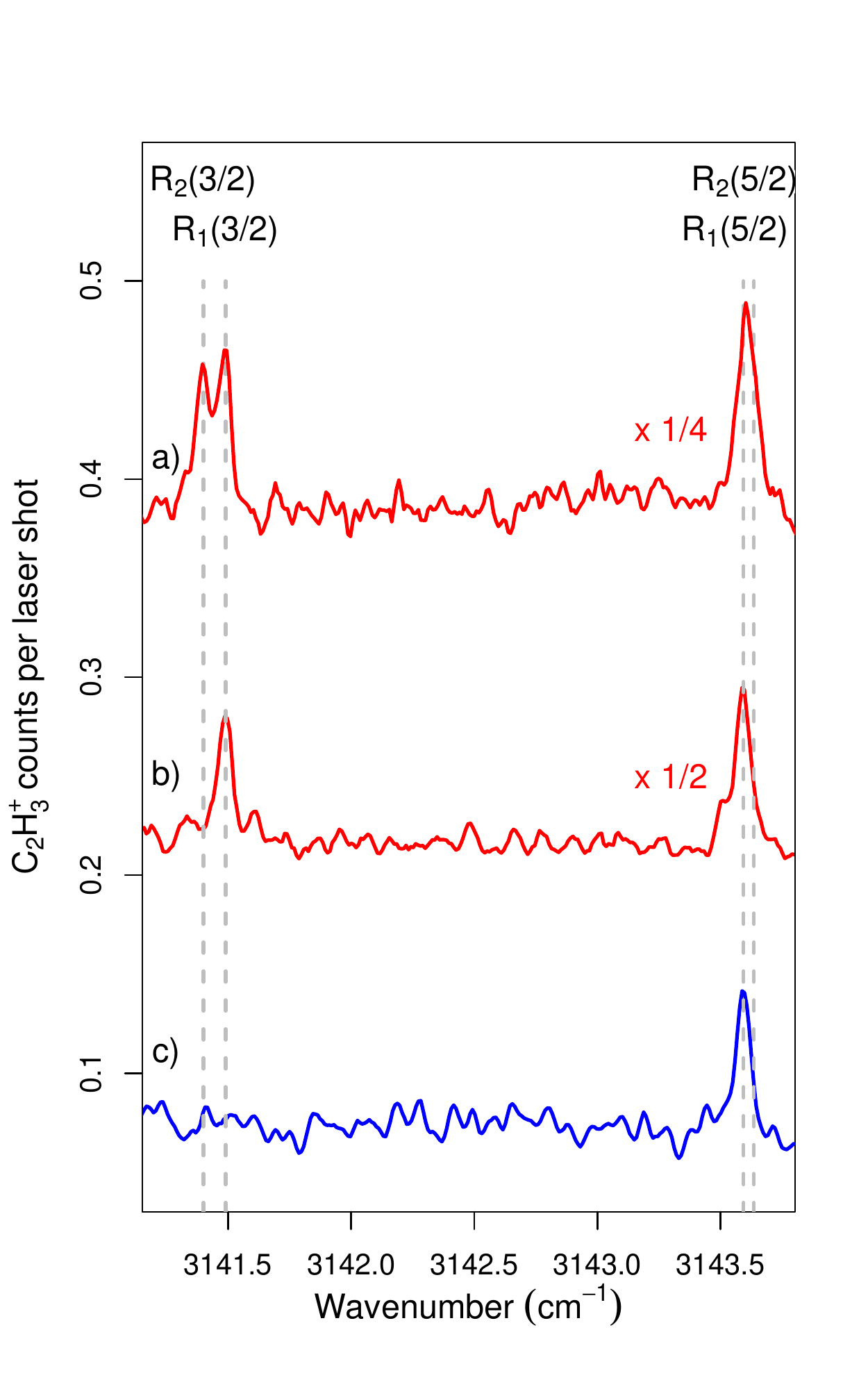}
\caption{IR spectra of the fundamental band of the asymmetric C-H stretching mode of C$_2$H$_2^+$. The C$_2$H$_2^+$ ions were prepared by photoionization above the $^2\Pi_{1/2}$ component (a), by photoionization at the position marked $\tilde{\nu}_1$ in Fig.~\ref{Figure3} (b), and following selection of the X$^+$ $^2\Pi_{3/2}$ ($v^+$ = 0, $J^+$ = 5/2) ionic level by MATI spectroscopy (c). Traces a), b), and c) were obtained by averaging over 1200, 2000, and 1200 laser excitation cycles, respectively.} 
\label{Figure4}
\end{figure}
Figure~\ref{Figure3} compares the MATI spectrum of the X$^+$ $^2\Pi_\textrm{u}$ ($v^+$=0) $\leftarrow$ X $^1\Sigma_\textrm{g}$ ($v$=0) ionizing transition of C$_2$H$_2$ (upper trace) with the high-resolution PFI-ZEKE photoelectron spectrum of the same transition reported by Rupper $et$ $al.$~\cite{Rupper2004}. Several lines of the MATI spectrum correspond to single transitions, such as the X$^+$ $^2\Pi_{3/2}$ ($J^+$ = 5/2) $\leftarrow$ X $^1\Sigma_\textrm{g}$ ($J^{\prime\prime}=1$) transition at the position $\tilde{\nu}_1=91954$\,cm$^{-1}$ (arrow marked $\tilde{\nu}_1$ in Fig.~\ref{Figure3}). At this spectral position, the pulse sequence used to form the ion beam guarantees that the ions in the beam are in the X$^+$ $^2\Pi_{3/2}$ ($J^+=5/2$) level prior to IR excitation. At the position $\tilde{\nu}_2=91982$\,cm$^{-1}$ (arrow marked $\tilde{\nu}_2$ in Fig.~\ref{Figure3}), the C$_2$H$_2^+$ ions are produced in the $J^+=3/2$ rotational level of the $^2\Pi_{1/2}$ spin-orbit component. By turning off the discrimination pulse, one can also generate an ion beam to which all ions contribute that can be generated by direct photoionization at the selected wavenumber.

These advantages are illustrated in Fig.~\ref{Figure4}. Trace a) was measured without discrimination pulse with the VUV wavenumber set at 92080\,cm$^{-1}$, $i.e.$, well above the X$^+$ $^2\Pi_{1/2}$ $\leftarrow$ X $^1\Sigma_\textrm{g}$ band. At this position all optically accessible rotational levels of the X$^+$ $^2\Pi_{3/2}$ and $^2\Pi_{1/2}$ spin-orbit components are generated. The spectral region displayed in this spectrum consists of R transitions from the $J^+=3/2$ and 5/2 levels of both spin-orbit components. When the VUV laser wavenumber is set at 91954\,cm$^{-1}$, $i.e.$, below the onset of the X$^+$ $^2\Pi_{1/2}$ $\leftarrow$ X $^1\Sigma_\textrm{g}$ band in the MATI spectrum (see Fig.~\ref{Figure3}), but still without discrimination pulse, the two R$_2$ lines disappear as can be seen in the middle trace (trace b)) in Fig.~\ref{Figure4}. The R$_1$(5/2) line is the only one remaining when the VUV laser is tuned to the position of  $\tilde{\nu}_1$ in Fig.~\ref{Figure3} and the discrimination pulse is switched on (trace c) in Fig.~\ref{Figure4}). This trace thus demonstrates full rotational state selectivity in the X$^+$ $^2\Pi_{3/2}$ spin-orbit component.

\begin{figure}[!ht]\centering
\includegraphics[width=1\linewidth]{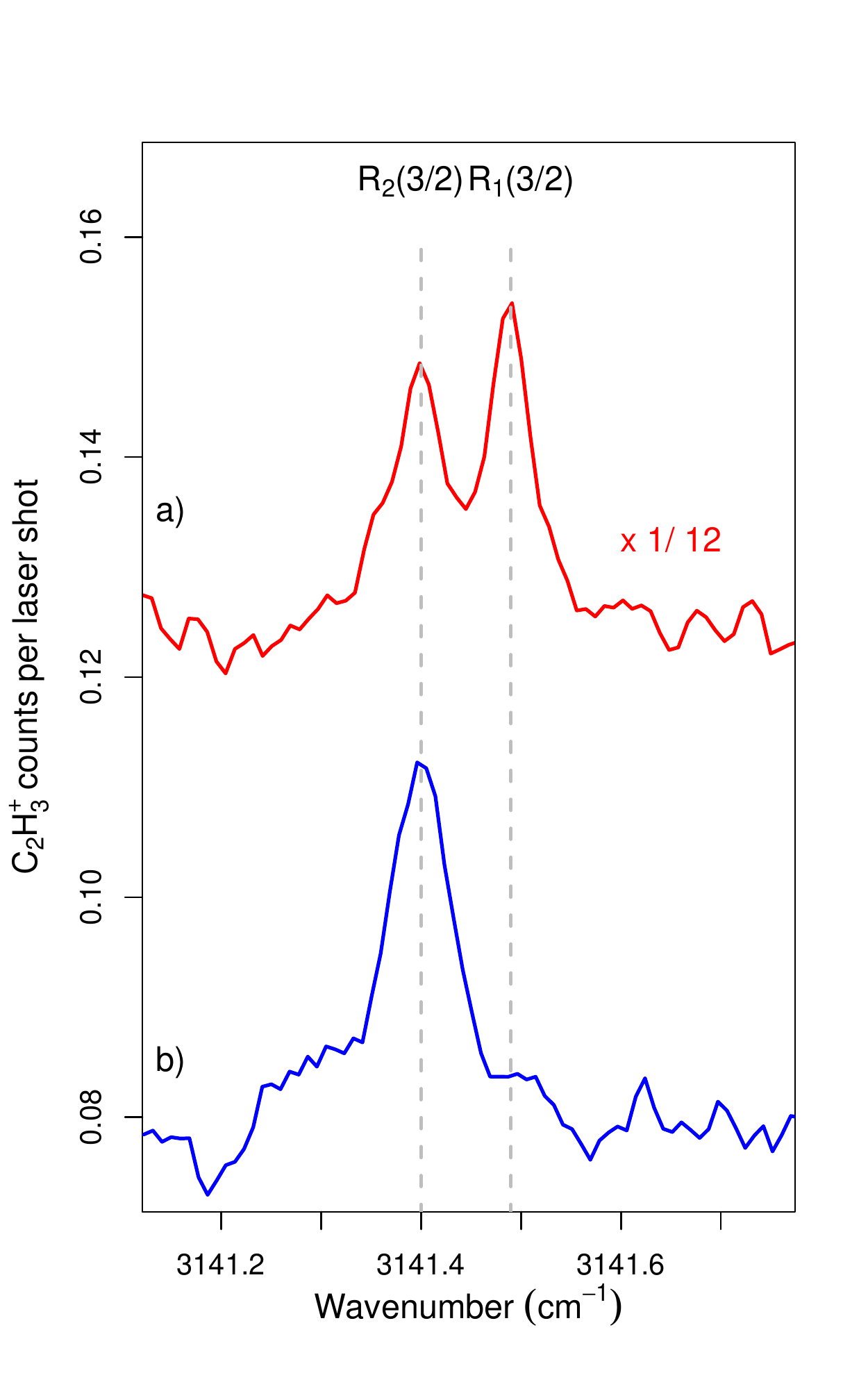}
\caption{IR spectra of the fundamental band of the asymmetric C-H stretching mode of C$_2$H$_2^+$. The C$_2$H$_2^+$ ions were prepared by photoionization above the $^2\Pi_{1/2}$ component (a) and following selection of the X$^+$ $^2\Pi_{1/2}$ ($v^+$ = 0, $J^+$ = 3/2) ionic level by MATI spectroscopy (b) at the position marked $\tilde{\nu}_2$ in Fig.~\ref{Figure3}.} 
\label{Figure5}
\end{figure}

Full rotational state selectivity can also be achieved in the upper ($^2\Pi_{1/2}$) spin-orbit component, as illustrated in Fig.~\ref{Figure5}. This figure compares the IR spectrum near the R$_{1,2}$(3/2) line pairs recorded with the VUV laser wavenumber fixed at the position marked $\tilde{\nu}_2$ in Fig.~\ref{Figure3} with (lower trace) and without (upper trace) discrimination pulse. This state selectivity is possible because the H$_2$ gas density in the reaction zone is low enough to ensure single-collision conditions and thus to avoid state redistribution by inelastic collisions prior to the reaction.

The lines observed in the IR spectrum presented in Figs.~\ref{Figure4} and \ref{Figure5} have a full width at half maximum of 0.035~cm$^{-1}$, limited by the bandwidth of the IR laser. In IR spectra recorded by applying the IR laser pulse before the field ionization pulse, we did not observe any frequency shift of the IR transitions but a slight broadening, which we attribute to the presence of the high-$n$ Rydberg electron. The transition wavenumbers, corrected for the estimated Doppler shift of $-$0.006~cm$^{-1}$, are compared in Table~\ref{tab} to the more precise transition wavenumbers reported in Ref.~[\onlinecite{Jagod1992}] with which they are in excellent agreement.

 \begin{table}[htb]
\centering
\begin{tabular}{c@{\hspace*{4mm}}c@{\hspace*{4mm}}c}
 & This work & Ref. [\onlinecite{Jagod1992}]  \\
\hline
R$_1$(3/2) & 3141.488(20) & 3141.491  \\
R$_2$(3/2) & 3141.400(20) & 3141.402 \\
R$_1$(5/2) & 3143.597(20) & 3143.594  \\
R$_2$(5/2) & 3143.629(30) & 3143.636 \\
\hline
\end{tabular}
\caption{Wavenumbers (in cm$^{-1}$) of the R$_{1,2}$(3/2) and R$_{1,2}$(5/2) transitions of the fundamental band of the asymmetric-stretching mode of C$_2$H$_2^+$ (X$^+$ $^2\Pi_\textrm{u}$). }
\label{tab}
\end{table}

\section{Conclusion}

We developed a method and built a new apparatus to record infrared spectra of mass- and state-selected cations. The method relies on the preparation of the individual vibrational and rotational state of molecular cations by photoionization and field-ionization methods and on the detection of the product of reactions induced by the absorption of infrared radiation. The state selectivity of the method reduces spectral congestion, which facilitates assignments. \textcolor{red}{The main obstacle to the general use of this method is the necessity to identify a reaction with the appropriate kinetics and thermodynamical properties.}

\begin{acknowledgments}
This work is supported financially by the Swiss National Science Foundation under project Nr. 200020-159848.

\end{acknowledgments}


\end{document}